\documentclass[twocolumn]{article}

\usepackage{amssymb,amsmath,latexsym,amsthm,amsfonts}
\usepackage{graphicx}
\usepackage{mathrsfs}
\usepackage{natbib}

\usepackage{color}
\newenvironment{theacknowledgments}
     {\section*{Acknowledgements}}
     {\par}

\bibpunct{(}{)}{,}{a}{}{,}

\title{{\bf On Bayesian Data Analysis}}

\author{{\sc Christian P.~Robert and Judith Rousseau}\thanks{C.P. Robert is Professor of Statistics at
Universit\'e Paris-Dauphine, CEREMADE, 75775 Paris cedex 16, and Head of the Statistics Lab at
CREST, INSEE, Malakoff, France.  Email: \texttt{xian@ceremade.dauphine.fr} Webpage:
\texttt{www.ceremade.dauphine.fr/$\sim$xian} Blog: \texttt{xianblog.wordpress.com}---Judith 
Rousseau is Professor of Statistics at Universit\'e Paris-Dauphine, CEREMADE, 75775 Paris cedex 16, and at
ENSAE, 92240 Malakoff, France.  Email: \texttt{rousseau@ceremade.dauphine.fr} Some of the quotes used in this Chapter
have been previously put to use in a debate about Bayesian statistics published as \cite{robert:2010}.} 
}

\begin{document}

\twocolumn 
\maketitle

\begin{abstract}
This introduction to Bayesian statistics presents the main concepts as well as the principal reasons advocated
in favour of a Bayesian modelling. We cover the various approaches to prior determination as well as the
basis asymptotic arguments in favour of using Bayes estimators. The testing aspects of Bayesian inference
are also examined in details.

\end{abstract}

\section{Introduction  : the Bayesian paradigm }

In this Chapter we give an overview of Bayesian data analysis, emphasising that it is {\em a method for summarising uncertainty and making 
estimates and predictions using probability statements conditional on observed data and an assumed model} 
\citep{gelman:2008}---which makes it valuable and useful in Statistics, Econometrics, and Biostatistics, among other fields.

We first describe the basic elements of Bayesian analysis. In the following, we refrain from embarking upon philosophical discussions
about the nature of knowledge \citep[][Chapter 10]{robert:2001} and the possibility of induction \citep{popper:miller:1983}, opting instead
for a mathematically sound presentation of a statistical methodology. We indeed believe that the most convincing arguments for adopting
a Bayesian version of data analyses are in the versatility of this tool and in the large range of existing applications.

\subsection{First principles}
Recall that, given a set of observations $x \in \mathcal X$, a statistical model is defined as a family of probability distributions on 
$\mathcal X$, say $(P_\theta, \theta \in \Theta)$ and the aim of statistical inference is to derive quantitative information 
about the unknown \textit{parameter} $\theta$. 
This information can be about explanatory features of the model, like the impact of the increase by one point of interest rates over inflation 
rate or the relevance of culling strategies during the latest foot-and-mouth epidemics in the UK
or yet the amount of cold dark matter in the Universe, or about predictive features, like the value of a particular stock the next day or the 
chances for a given individual of catching the H5N1 flu over the coming three mouths. Inference is quantitative in that it provides 
numerical values for the quantities of interest and numerical evaluations of the uncertainty surrounding those values as well.

Since all models are approximations of the real World, the choice of a sampling model is wide-open for criticisms:
{\em Bayesians promote the idea that a multiplicity of parameters can be handled via hierarchical, typically exchangeable, 
models}
\citep{gelman:2008}. This is however a type of criticism that goes far beyond Bayesian modelling and 
questions the relevance of completely built models for drawing inference or running predictions.

The central idea behind Bayesian modelling is that the uncertainty on the unknown parameter $\theta$ is  
better modelled as randomness and consequently a probability distribution $\Pi$ is constructed on 
$\Theta$.  In particular $P_\theta$  then represents the probability distribution of the observation $x$ given that the parameter is equal to $\theta$, i.e. the conditional probability distribution of $x$ given $\theta$. If $\Pi$ is a probability on $\Theta$, with density $\pi$ with respect to some measure $\nu$ on $\Theta$, then we can define a joint distribution for the observation and the parameter $(x,\theta)$
$$P_\pi((x,\theta)\in A\times B) = \int_{\theta \in B} P_\theta(A)\pi(\theta)\text{d}\nu(\theta).$$
 For the sake of simplicity we consider only models $(P_\theta, \theta\in \Theta)$ that allow for a dominating measure, $\mu$ (say the Lebesgue measure), and we denote by $f(.|\theta)$ the density of $P_\theta$  with respect to $\mu$ (the likelihood). Then the joint distribution of $(x,\theta)$ has density 
 \begin{eqnarray}\label{def:joint}
 p_\pi(x,\theta) = f(x|\theta)\pi(\theta), 
 \end{eqnarray}
with respect to $\mu\times \nu$. Using Bayes theorem we can define the distribution of the parameter $\theta$ 
given the observations by its density with respect to $\nu$:
\begin{eqnarray}\label{def:posterior}
\pi(\theta|x) &=& \frac{ f(x|\theta)\pi(\theta) }{ \int_\Theta f(x|\theta)\pi(\theta)\text{d}\nu(\theta) },
\end{eqnarray}
and denote the denominator by 
$$
 m(x) = \int_\Theta f(x|\theta)\pi(\theta)\text{d}\nu(\theta)\,.
$$  
The probability $\Pi$ ($\pi$, respectively) on $\Theta$ is called the \textit{prior distribution } (density, respectively),  
the conditional probability (\ref{def:posterior}) of $\theta$ given $x$ is called the \textit{posterior 
distribution }(density, respectively)  and $m(x)$ is the \textit{marginal density} of the observation $x$. Then, Bayesian 
analysis is based entirely on the posterior distribution (\ref{def:posterior}), for all inferential purposes,
e.g.~to draw conclusions on the parameter $\theta$ or on some functions of the parameter $\theta$, to make predictions, 
to test the plausibility of a hypothesis or to check the fit of the model. 

There are many arguments which make such an approach compelling. Without entering into philosophical and 
epistemological arguments on the nature of Science \citep{jeffreys:1939,mackay:2002,jaynes:2003}, 
we briefly state what we view as the main practical appealing 
features of introducing a prior probability on $\theta$. First such an approach allows to incorporate prior 
information in a natural way in the model, as explained in Section \ref{sec:witch}; second, by defining a probability 
measure on the parameter space $\Theta$, the Bayesian approach gives a proper meaning to notions such as \textit{the 
probability that }$\theta$ \textit{belongs to a specific region} which are particularly relevant when constructing 
measures of uncertainty like confidence regions or when testing hypotheses. Furthermore, the posterior distribution 
\eqref{def:posterior} can be interpreted as the actualisation of the knowledge (uncertainty) on the parameter after observing 
the data. We stress that the Bayesian paradigm does not state that the model within
which it operates is the ``truth", no more that it believes that the corresponding prior distribution it requires has a connection
with the ``true" production of parameters (since there may even be no parameter at all). It simply provides an inferential machine that
has strong optimality properties under the right model and that can similarly be evaluated under any other well-defined alternative model.  
Furthermore, the Bayesian approach is such that {\em techniques allow prior beliefs to be tested and discarded 
as appropriate} \citep{gelman:2008}, in agreement that 
the overall principle that a {\em Bayesian data analysis has three stages: formulating a model, fitting the model 
to data, and checking the model fit} \citep{gelman:2008}, so there seems to be little reason for not using a given model at an
earlier stage even when dismissing it as ``un-true" later (always in favour of another model).

In the above formulation, note that $\Theta$ can be endowed with quite different features: it
can be a finite dimensional set (as in parametric models), an infinite dimensional set (as in most semi/non parametric models) 
or a collection of various sets with no fixed dimension (as in model choice).

As an example, consider the following contingency table on survival rate for breast-cancer patients
with or without malignant tumours, extracted from \cite{bishop:fienberg:holland:1975}, the ultimate goal
being to distinguish between both types of tumour in terms of survival probability:
\begin{verbatim}
                  survival
    age  malignant yes no 
under 50        no  77 10
               yes  51 13
   50-69        no  51 11
               yes  38 20
above 70        no   7  3
               yes   6  3
\end{verbatim}
\begin{figure}[hbp]
\includegraphics[width=.5\textwidth]{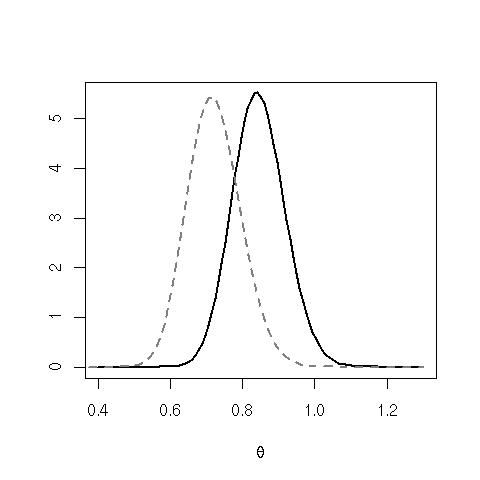}
\caption{\label{fig:cancer}
Representation of two gamma posterior distributions differentiating between malignant {\em (dashes)}
versus non-malignant {\em (full)} breast cancer survival rates.}
\end{figure}
Then if we assume that both groups (malignant versus non-malignant) of survivors are Poisson distributed $\mathcal{P}(N_t\theta)$,
where $N_t$ is the total number of patients in this age group, i.e.
$$
f(x_t|\theta,N_t) = e^{-\theta N_t}\frac{ (\theta N_t)^x_t}{x_t!}, \quad x\in \mathbb{N}\,,
$$
then we obtain a likelihood
$$
L(\theta|D) = \prod_{t=1}^3 (\theta N_t)^{x_t} \exp\{-\theta N_t\}
$$
which, under an exponential $\theta\sim\mathcal{E}xp(2)$ prior---whose rate $2$ is chosen here for 
illustration purposes---, leads to the posterior
$$
\pi(\theta|D)\propto \theta^{x_1+x_2+x_3} \exp\left\{-\theta(2+N_1+N_2+N_3)\right\}
$$
i.e.~a Gamma $\Gamma(x_1+x_2+x_3+1,2+N_1+N_2+N_3)$ distribution. The choice of the exponential parameter corresponds to a $50\%$
survival probability. In the case of the non-malignant breast cancers, the parameters of
the Gamma distribution are $a=136$ and $b=161$, while, for the malignant cancers, they are $a=96$ and $b=133$. Figure \ref{fig:cancer}
shows the difference between those posteriors.

\subsection{Extension to improper priors}

In many situations, it is useful to extend the above setup to prior
measures that are not probability distributions but $\sigma$-finite measures with infinite mass, i.e.
$$\int_\Theta \pi(\theta)\text{d}\nu(\theta) = +\infty,$$
since, provided that
\begin{equation}\label{eq:proper}
\int_\Theta f(x|\theta)\pi(\theta)\text{d}\nu(\theta) <+\infty,
\end{equation}
almost everywhere (in $x$), the quantity  (\ref{def:posterior}) is still well-defined as a probability 
density as when using a regular posterior probability as prior \citep{hartigan:1983,berger:1985,robert:2001}. 
Such extensions are justified for a variety of reasons, ranging from topological coherence---limits of  
Bayesian procedures often partake of their optimality properties \citep{wald:1950} and should therefore be included 
in the range of possible procedures---to robustness---a measure with an infinite mass is much more robust than a 
true probability distribution with a large variance---and improper priors are typically encountered in situations where there 
is little or no prior information, inducing flat, i.e.~uniform, distributions on the parameter space (or on some transforms of 
the parameter space). Indeed it is quite common, for complex models, to have little or no information on some of the parameters 
present in the model and using improper priors for such parameters has many advantages. Note however that in such cases the 
marginal density $m(x)$ does not define a probability on $\mathcal X$ (and that the existence condition \eqref{eq:proper} needs to be
checked). This drawback has an importance consequence for Bayesian model comparison as explained in Section \ref{seq:test}.

Note also that some improper priors never allow for well-defined posteriors, no matter how many observations there are in the
sample. One such example is when the prior is $\pi(\theta) = \exp(+\theta^2)$ and the observations are iid Cauchy. Another and less
anecdotic example occurs in mixture models, under exchangeable improper priors on the components \citep{lee:marin:mengersen:robert:2008}.

\subsection{Bayesian decision theory}

As a general {\em modus vivendi}, let us first stress that inference as a whole is meaningless unless it is evaluated.
The evaluation of a statistical procedure, i.e. determining how well or how bad the inference performs, requires the definition 
of a comparison criterion, called a loss function. Set $\mathcal D$  the set of all possible results of the inference (corresponding
to the decision set in game theory). An estimator 
is then a function from $\mathcal X$ into $\mathcal D$. (With an obvious abuse of notation, we will also use
$\mathcal D$ for the set of estimators.) For instance, the aim is to estimate $\theta$, then
$\mathcal D =\Theta$; if the  aim is to test for some hypothesis, then $\mathcal D = \{0,1\}$,
and $\mathcal D = \mathcal X_1$ the set of a future observation if the aim is to predict a future observation. 
A  loss function $L$ is a function on 
$\Theta \times \mathcal D$, expressing what the loss (cost) is for considering a decision $\delta$ when $\theta$ is the 
\textit{true} value. Typical (formal) loss functions used for estimation and test
are quadratic losses ($L(\theta, \delta) = |\!| \theta -\delta|\!|^2$)
and 0-1 losses (1 if decision is wrong, 0 if it is right), respectively. 
Other loss functions can (should) be constructed, depending of the problem
at hand, and they are strongly related to the notion of utility function encountered in economy and game theory \citep{berger:1985}. 

Given a statistical model $(P_\theta, \theta\in \Theta)$ on $x\in \mathcal X$, a  prior $\pi$ on $\theta \in \Theta$ 
and a loss function $L)$, the (optimal) Bayesian procedure (estimator)  is then defined as the 
decision function $\delta$ minimising the integrated risk $r(\pi,\delta)$:
$$
\delta^\pi = \mbox{argmin}_{\delta \in \mathcal D} r(\pi,\delta)
$$
where
$$
 r(\pi,\delta) = \int_{\Theta \times \mathcal X} L(\theta, \delta(x)) f(x|\theta)\pi(\theta)\text{d}\mu(x)\text{d}\nu(\theta).
$$
Such a procedure is called a Bayes estimator. 
Using the fact \citep{robert:2001}, that such estimators can be computed pointwise
as minimising the posterior risk: $\forall x \in \mathcal X$, 
\begin{eqnarray*}
\delta^\pi(x) 
 & =& \mbox{argmin}_{\delta \in \mathcal D} \int_{\Theta} L(\theta, \delta(x))\pi(\theta|x) \text{d}\nu(\theta),
\end{eqnarray*}
it is possible to derive explicit expression of Bayes estimates for many common loss functions. In particular, 
the Bayes estimator associated with the quadratic loss and the posterior distribution $\pi(.|x)$ is the  posterior mean
$$
\delta^\pi(x) = \int_\Theta \theta \pi(\theta|x)\,\text{d}\theta\,.
$$

Note that the integrated risk $r(\pi,\delta)$ can also be expressed as $\int_\Theta R(\theta,\delta)
\pi(\theta)\text{d}\nu(\theta)$, where $R(\theta,\delta) = \int_{\mathcal X }L(\theta,\delta(x))f(x|\theta)\text{d}\mu(x)$ 
is the frequentist risk, so that Bayes estimates are also often optimal in the frequentist sense. (It can be shown in 
particular that any admissible estimator is the limit of Bayes estimators, see \citealp{berger:1985} or \citealp{robert:2001}).

\section{On the selection of the prior}\label{sec:witch}

A critical aspect 
is the determination of the prior 
distribution $\pi$ and its clear influence on the inference. It is straightforward to come up with examples where 
a particular choice of the
prior leads to absurd decisions. Hence, for a Bayesian analysis to be sound the prior distribution needs to be well-justified. 
Before entering into a brief description of the various ways of constructing prior distributions, note that as part of model 
checking, it is necessary in every Bayesian analysis to assess the influence of the choice of the prior, for instance through 
a sensitivity analysis.  Since the prior distribution models the knowledge (or uncertainty) prior to the observation of 
the data, the sparser the prior information is, the flatter the prior should be. The advantage of incorporating prior information via 
a prior distribution is rather universally accepted and we therefore first describe ways of eliciting prior distributions from prior 
knowledge.

\subsection{Elicited priors}
The elicitation of prior distributions from prior knowledge consists in the construction of the prior probability $\pi(\theta)$ 
using all items of prior information available to the modeller. This prior information may come from expert opinions or from bibliographic 
data or yet from earlier analyses, as in meta-analysis. There exists a vast literature on prior elicitation based on expert opinions, 
which is a much more complex process than is usually acknowledged in most Bayesian statistical notebooks, see Section 2 of this book
for a more complete discussion on prior elicitation based on expert opinions.

In particular the prior information is rarely rich enough to entirely define  a prior distribution, therefore it is customary
to choose a prior distribution within a parametric class of possible distributions: $\pi(\theta|\gamma)$, where $\gamma 
\in \Gamma$ is called a hyperparameter. In such cases the prior information is summarised through the choice of $\gamma$. For instance,
\citet{albert:grenier:denis:rousseau:2008} use bibliographic prior information to construct a prior distribution on the probability 
of cross-contamination from a contaminated broiler in a household, say $p$. The prior distribution of $p$ is assumed to 
be a Beta $\mathcal{B}e(a,b)$  distribution, 
$$
\pi(p|a,b) \propto p^{a-1} (1-p)^{b-1}\,,\quad 0<p<1\,,
$$
and the parameters $(a,b)$ of the Beta distribution are assessed using two cross--contamination models in the 
literature which lead to a probability of transfer between $1/3$ and $2/3$, which was translated into a $\mbox{Beta}(8,8)$ prior on 
$p$, as it corresponds to a prior mean of $0.5$ and to a $95\%$ prior confidence interval equal to $(0.27,0.73)$. 
See also \cite{dupuis:1995b} for an example of expert elicitation of the Beta parameters on some capture and survival probabilities 
in a lizard population, or the Chapter of B\"ocker, Crimmi and Fink in this volume where beta priors are elicited 
to model correlations between risk types.

\subsection{Conjugate priors} 
Among the possible parametric families $\pi(\theta|\gamma)$, $\gamma \in \Gamma$, conjugate priors form appealing parametric families, 
merely for computational reasons \citep{berger:1985,robert:2001}. A family of distribution prior distributions $\pi(\theta|\gamma)$, 
is said to be conjugate to the likelihood $f(x|\theta)$ if the posterior also belongs to the same family, i.e. when the prior is equal to 
$\pi(\theta|\gamma_0)$ then there exists a $\gamma(x, \gamma_0)\in \Gamma$ such that the posterior is equal to $\pi(\theta|\gamma(x, 
\gamma_0))$. The actualisation of the information due to observing the data $x$ is then modelled as a change 
of hyperparameter from $\gamma_0$ to 
$\gamma(x,\gamma_0)$. Exponential families (as models for the observation $x$) are almost in one-to-one correspondence with
sampling distributions allowing for conjugate priors. As an example, 
\cite{carlin:louis:2001} consider an observed random variable $X$ 
that is the number of pregnant women arriving at a given hospital to deliver their babies within a given month, 
which they model as a Poisson $\mathcal{P}(\theta)$ distribution with parameter $\theta>0$.
A conjugate family of priors for the Poisson model is the collection of gamma distributions $\Gamma(a,b)$, since
$$
f(x|\theta) \pi(\theta|a,b) \propto \theta^{a-1+x} e^{-(b+1)\theta} 
$$
leads to the posterior distribution of $\theta$ given $X=x$ being the gamma distribution $\Gamma(a+x,b+1)$. The computation of estimators, of
confidence regions---called credible regions within the Bayesian literature to distinguish the fact that those regions are evaluated on 
the parameter space rather than on the observation space \citep{berger:1985}---or of other types of summaries of interest on the posterior 
distribution then becomes straightforward. For instance in the above Poisson--Gamma example, the Bayesian estimator of the average number 
of arrivals, associated with the quadratic loss, is given by $\hat{\theta} = (a+x)/(b+1)$, the posterior mean. 

The apparent simplicity of conjugate priors should however not make them excessively appealing, since there is no strong justification 
to their use. One of the difficulties with such families of priors is the influence of the hyperparameter $\gamma_0$. If the prior 
information is not rich enough to justify a specific value of $\gamma$, fixing $\gamma = \gamma_0$ arbitrarily is problematic, 
since it does not take into account the prior uncertainty on $\gamma_0$ itself. To improve on this aspect of conjugate priors, 
a usual fix is to consider a \textit{hierarchical prior}, i.e.~to assume that $\gamma $ itself is random and to consider 
a probability distribution with density $q$ on $\gamma$, leading to
  \begin{eqnarray*}
  \theta|\gamma &\sim& \pi(\theta|\gamma)\\
  \gamma & \sim& q(\gamma)\,,
  \end{eqnarray*}
as a joint prior on $\theta,\gamma)$. The above  is equivalent to considering, as a prior on $\theta$
$$
\pi(\theta) = \int_\Gamma \pi(\theta|\gamma)q(\gamma)\text{d}\gamma\,.
$$ 
Obviously $q$ may also depend on some hyperparameters $\eta$. Higher order levels in the hierarchy are thus possible, 
even though the influence of the hyper(-hyper-)parameter $\eta$ on the posterior distribution of $\theta$ is 
usually smaller than that of $\gamma$. But multiple levels are nonetheless useful in complex populations as those
found in animal breeding \citep{sorensen:gianola:2002}.

In many applications prior information  is quite vague or at least vague enough on some parts of the model, 
in which case it is important to derive priors that have desirable properties and that are as little arbitrary 
or subjective as possible. Such constructions are commonly called {\em non informative}. While this denomination 
is misleading, and should be replaced by the less judgemental {\em reference prior} denomination, 
we nonetheless follow suit and use it in the following subsections, since it is the most common denomination found
in the literature \citep{kass:wasserman:1996}. 
  
 \subsection{Non informative priors}

Non informative priors are expected to be flat distributions, possibly improper. An apparently natural way of constructing such priors would 
be to consider a uniform prior, however this solution has many drawbacks, the worst one being that it is not invariant under a change of 
parameterisation. To understand this consider the example of a Binomial model: the observation $x$ is a $\mathcal B(n,p)$ random variable, 
with $p \in (0,1)$ unknown. The uniform prior $\pi(p) = 1$ could then sound like the most natural non informative 
choice; however, if, instead of the mean parameterisation by $p$, one considers the logistic parameterisation $\theta = \log (p/(1-p))$ 
then the uniform prior on $p$ is transformed into the logistic density
$$
\pi(\theta) = e^\theta / (1 + e^\theta)^2
$$ 
by the Jacobian transform, which obviously is not uniform. 

To circumvent this 
lack of invariance per reparameterisation, \cite{jeffreys:1939} proposed the following choice now known as {\em Jeffreys' prior}
\begin{eqnarray} \label{jeffreys}
\pi(\theta) &\propto& \sqrt{|i(\theta)|}, 
\end{eqnarray}
where $i(\theta)$ is the Fisher-information matrix and $|i(\theta)|$ denotes its determinant. The above construction 
is obviously invariant per reparameterisation and has many other interesting features specially in one-dimensional setups
(see \citealp{robert:chopin:rousseau:2009} for a reassessment of Jeffreys' impact on Bayesian statistics). 
In particular, in the one-dimensional parameter case, the Jeffreys prior is also the matching prior (see \citealp{robert:2001}, 
Chapters 3 and 8), and the reference prior defined by Bernardo \citep{bernardo:1979, clarke:barron:1990}.  For instance, when
$P_\theta$ is a location family, i.e.~when $f(x|\theta)=g(x-\theta)$, the Fisher information is constant and thus the Jeffreys prior
is $\pi(\theta)=1$. Note that in many cases like the above the Jeffreys prior is improper.

In multivariate setups, Jeffreys' construction is not so well-justified and 
it may lead to not-so-well-behaved priors. A famous example is the Neyman--Scott problem where two groups of observations are such 
that in each group all observations are distributed from $x_{i,j} \sim \mathcal N(\mu_i, \sigma^2)$, $i=1,...,n$, $j=1,2$. 
In this case Jeffreys' prior is given by $\pi(\mu_1,...,\mu_n, \sigma) \propto \sigma^{-(n+1)}$, and the Bayes estimator 
of $\sigma^2$ associated with the quadratic loss function is equal to 
$$
\mathbb{E}^\pi\left[ \sigma^2|x_{1,1},...,x_{n,2}\right] = \sum_{i=1}^n \frac{ (x_{i,1}- x_{i,2})^2 }{ 4(n-1)} \,,
$$
which converges to $\sigma^2/2$ as $n$ goes to infinity, thus leading to an inconsistent estimate. Although this seems like an 
artificial example it is actually of wider interest, since in the normal linear regression model Jeffreys' prior is 
proportional to $\sigma^{-p-2}$ where $p$ is the number of covariates. This dependence on $p$ makes it rather unappealing,
even though the alternative $g$-prior of \cite{zellner:1986} discussed below suffers from the same drawback. Another standard
example discussed further in Section \ref{sec:nuisance} is when estimating $||\theta||^2$ when $\theta$ is the $n$-dimensional 
mean of an $n$-dimensional normal vector.

The ultimate attempt to define a non informative prior is in our opinion
Bernardo's (\citeyear{bernardo:1979}) definition through the information theoretical device of Kullback divergence (see also 
\citealp{berger:bernardo:1992} or \citealp{berger:bernardo:sun:2009}). The idea is to split the parameter into groups say $(\theta_{(1)},...,\theta_{(p)})$ where $\theta_{(1)}$ is more interesting than $\theta_{(2)}$, which is more interesting than $\theta_{(3)}$ and so on. This can be seen as a generalisation of the usual splitting into a  parameter of interest and a nuisance parameter. Then the Bernardo's reference prior is constructed iteratively as some sorts of Jeffreys' priors in each of the submodels, see also \cite{robert:2001} 
for a more precise description of the iterative construction.  Quite obviously, this is not the unique possible approach, it depends on a choice of
information measure, does not always lead to a solution, requires an ordering of the model parameters that involves some prior information
(or some subjective choice) but, as long as we do not {\em think of those reference priors as representing ignorance}  \citep{lindley:1973},
they can indeed be {\em taken as reference priors, upon which everyone could fall back when the prior information is missing}
\citep{kass:wasserman:1996}.

\subsection{Some asymptotic results}

A well-known phenomenon is the decrease of influence of the prior as the sample size (or the information in the data) increases. 
We shall recall here these results in the simpler case of i.i.d observations, however these results can be extended to non i.i.d.~cases 
such as dependent observations under stationary and mixing properties, Gaussian processes and so on. Generally speaking in most parametric 
cases, the posterior distribution concentrates towards the true parameter value as $n$ goes to infinity so that posterior estimates will 
converge to the true values, as $n$ goes to infinity. This first type of results ensures that point estimates are satisfactory, as far as
asymptotic convergence is concerned. 

Another important aspect of the asymptotic analysis of Bayesian procedures is to understand how 
the measures of uncertainty derived from the posterior can be related to frequentist measures of 
uncertainty. Such a relation can be deduced from the Bernstein Von Mises property, which can be stated in the following way:
Assume that the vector of observations $x = (x_1,...,x_n):= x^n$ is made of i.i.d observations from a distribution $f(.|\theta)$, which 
is regular, see for instance \cite{ghosh:ramamo:2003} for more precise conditions, and let $\pi$ be a prior density, which is positive and 
continuous on $\Theta$, then the posterior distribution can be approximated in the following way, when $n$ goes to infinity: 
for all $A \subset \Theta$
\begin{eqnarray*}
P^\pi\left[ \sqrt{n}(\theta-\hat{\theta}) \in A|x^n\right] \approx P\left[ \mathcal N(0, i_1(\hat{\theta})^{-1}) \in A\right],
\end{eqnarray*}
where $\hat{\theta}$ is the maximum likelihood estimator and $i_1(\hat{\theta})$ is the Fisher information matrix per observation 
calculated at $\theta = \hat{\theta}$. In other words the posterior distribution resembles a Gaussian distribution centred at 
$\hat{\theta}$ with covariance matrix $i^{-1}(\hat{\theta})/n$, when $n$ is large. 

This result has many interesting implications. 
The first consequence is that, to first order, the influence of the prior disappears as $n$ goes to infinity. It also allows for 
quick approximate computations in the case of large samples, and it implies that to first order Bayesian and frequentist inference 
(based on the likelihood) essentially give the same answers. Although devising procedures giving the same answers as frequentist 
procedures is not an ultimate aim of the Bayesian analysis, it is of importance to ensure that Bayesian procedures ultimately have 
also good frequentist properties. The asymptotic equivalence between the Bayesian and the frequentists answers (to first order) 
hold in wide generality for finite dimensional models. When the dimension of the parameter grows with the number of observations 
or is infinite, then this is often not true anymore, see for instance \cite{freedman:1999} and \cite{rivoirard:rousseau:2009}.

Although these asymptotic results have a strong frequentist flavour, in the sense that they are obtained by assuming that there is 
a fixed true parameter $\theta_0$ and as new data comes in the posterior concentrates around the true parameter like a Gaussian 
distribution, they are also appealing from the  subjectivists points of view where {\em probabilities represent degrees of belief 
and there are no objective  probability model}, see \cite{diaconis:freedman:1986} for a more 
precise discussion on this issue.

\section{Measures of uncertainty: credible regions}

Recall that the whole inference about $\theta$ is deduced from the posterior distribution, $\pi(\theta|x)$, 
including estimates as major summaries, but the posterior distribution gives us much more information than simply point estimates. 
In particular, different measures of uncertainty can be derived from the posterior and among the various measures credible regions 
are the most popular. A set $C \subset \Theta$ is an $\alpha$ - credible region if and only if 
\begin{equation}\label{eq:edible}
P^\pi\left[ \theta \in C| x\right] \geq 1 -\alpha.
\end{equation}
Contrariwise to frequentist confidence regions, the notion of coverage probability is directly understood as a probability on 
$\theta$ and is therefore straightforward to interpret. Among all credible regions defined by \eqref{eq:edible}, those having minimal volume 
are particularly interesting. It turns out, see \cite{robert:2001}, that they are defined as highest posterior density (HPD) regions:
\begin{eqnarray*}
C_\alpha^\pi = \{ \theta; \pi(\theta)f(x|\theta) \geq k_\alpha(x)\} 
\end{eqnarray*}
where $ k_\alpha(x)$ is the largest value such that
\begin{eqnarray*}
P^\pi\left[  \theta \in C_\alpha^\pi|x\right] \geq 1-\alpha.
\end{eqnarray*}
(Note that we define the bound $k_\alpha(x)$ in terms of the product prior$\times$likelihood in order to bypass the difficulty with
the normalising constant $m(x)$.)

\begin{figure}[hbp]
\includegraphics[width=.5\textwidth]{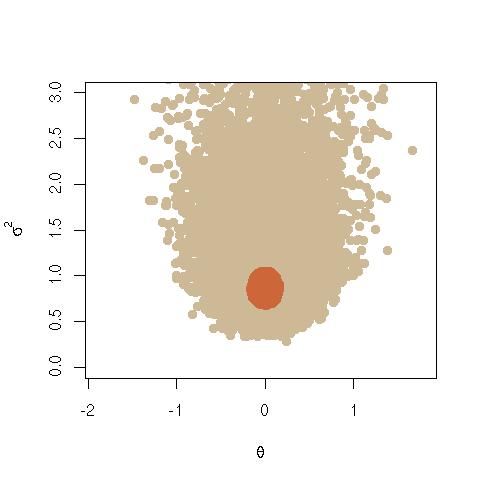}
\caption{\label{fig:hpd}
Representation of a Gibbs sample of $10^3$ values of $(\theta,\sigma^2)$ for the normal model,
$x_1,\ldots,x_n\sim\mathcal{N}(\theta,\sigma^2)$ with
$\overline x=0$, $s^2=1$ and $n=10$, under Jeffreys' prior,
along with the pointwise approximation to the $10\% $ HPD region {\em (in darker hues)} ({\em Source:} \citealp{robert:wraith:2009}).}
\end{figure}

Although the analytic determination of $k_\alpha(x)$ is often challenging, the approximation of this bound based on a sample from
$\pi(\theta|x)$, $\theta^{(1)},\ldots,\theta^{(p)}$, can be easily derived from an ordering of the values 
$\pi(\theta^{(i)})f(x|\theta^{(i)})$ as the corresponding $(1-\alpha)$-th quantile. For instance, if a Poisson $X\sim\mathcal{P}(\theta)$ count
is associated with a Gamma $\Gamma(a,b)$ prior, the posterior $\Gamma(a+x,b+1)$ leads to the HPD region
$$
\{\theta; \theta^{a+x-1}\,\exp(-(b+1)\theta) \ge k_\alpha(x)\}
$$
whose determination requires a numerical construct. On the other hand, if a sample $\theta^{(1)},\ldots,\theta^{(p)}$ 
from the posterior $\Gamma(a+x,b+1)$ is available, then the HPD bound $k_\alpha(x)$ can be estimated as the
$(1-\alpha)$-th quantile of the values $[\theta^{(i)}]^{a+x-1}\,\exp(-(b+1)\theta^{(i)})$'s. Figure \ref{fig:hpd}
illustrates a similar derivation in the case of a normal $\mathcal{N}(\theta,\sigma^2)$ model with both parameters unknown.

Credible regions have nice interpretations and are optimal under a volume criterion, 
as Bayesian estimators of the confidence sets $C$. In a wide generality, they further attain good frequentist coverage in the sense that
$\mathbb{P}_\theta(\theta\in C)=1-\alpha+\text{O}(n^{-1/2})$ for most prior distributions $\pi$,
where $n$ denotes the sample size \citep[][Chapter 5]{welch:peers:1963,robert:2001}. Credible regions however suffer 
from a lack of invariance to changes of parameterisation, i.e. if $\theta$ is a given parameterisation of interest and $C_\alpha^\pi$ 
is the HPD region constructed as above, then if $\eta = g(\theta)$ is another parameterisation, 
$g( C_\alpha^\pi ) = \{ \eta = g(\theta); \theta \in C_\alpha^\pi \}$ is not necessarily the HPD 
region for the $\eta$ parameterisation (see \citealp{druilhet:marin:2007} for a detailed analysis of this
phenomenon).

\section{Nuisance parameters : integrated likelihood} \label{sec:nuisance}

In many applied problems, one is only interested in some components of the parameter, the remaining part of the parameter 
being then called the nuisance parameter. This distinction opposes the parameter of interest, say $\psi$
within $\theta = (\psi, \lambda)$, where $\psi $ is the parameter of interest and $\lambda$ is the nuisance parameter. 
Dealing with nuisance parameters is quite problematic in a frequentist framework, whether one is interested in parameter estimation, 
in confidence regions determination or in testing. Likelihood approaches need to define \textit{proper} likelihoods for $\psi$, 
which in complete generality is not possible. Hence, they use approximations and modifications of proper likelihoods such as 
partial likelihoods or modified profile likelihoods, see \cite{severini:2000} for a more complete discussion on these issues. 

On the opposite, the Bayesian framework offers is a most natural way of dealing with nuisance parameters and for defining proper 
profile likelihoods : integrating out the nuisance parameter. In other words the Bayesian marginal likelihood 
for $\psi$ under the prior $\pi(\lambda | \psi)$ is given by
\begin{eqnarray} \label{margin:likeli}
f_\pi(x|\psi) &=& \int_{\lambda} f(x|\psi,\lambda) \,\text{d}\pi(\lambda|\psi).
\end{eqnarray}
This approach offers many advantages: (1) If the conditional prior $\pi(\lambda|\psi)$ is proper, then $f_\pi(x|\psi)$ 
as defined in (\ref{margin:likeli}) is  a proper likelihood, in the sense that it is the density of $x$ under some model 
parameterised by $\psi$ alone; (2) Integrating $\lambda$ out implicitly takes into account the uncertainty on $\lambda$, 
contrary to the profile likelihood, or to any other kind of plug-in likelihood defined by $f(x|\psi, \hat{\lambda}_\psi)$, 
where $\hat{\lambda}_\psi$ is some \textit{estimate} of $\lambda$ given $\psi$. In particular uncertainty measures derived 
from $f_\pi(x|\psi)$ are not biased downwards due to the replacement of $\lambda$ by $\hat{\lambda}_\psi$. Hence there 
is no need to correct further for this uncertainty, which is usually necessary when dealing with plug-in likelihoods, 
leading to penalised likelihoods. This is of particular interest in model selection, when the parameter of interest 
is the model itself, as discussed in Section \ref{seq:test}.

However, if $\pi(\lambda|\psi)$ is an improper prior, then $f_\pi(x|\psi)$ is not necessarily a \textit{likelihood}, in particular 
$\int_{\mathcal X} f_\pi(x|\theta)\,\text{d}x = +\infty$ may occur. A well-known example of such misbehaviour is the case of the
so-called marginalisation paradoxes, see for instance Robert (2001, Chapter 3). As another example of badly behaved 
marginal likelihood, consider the case presented in Robert (2001, Chapter 3) and \cite{liseo:2006} where 
the observations $x_i \sim \mathcal N(\mu_i, 1)$, $i=1,...,p$, are independent and where the parameter of interest is $\psi = 
|\!| \mu|\!|^2 /p = \sum_{i=1}^p\mu_i^2/p$ where $\mu =(\mu_1,...,\mu_p)$ and the nuisance parameter is $\lambda = \mu/|\!| 
\mu |\!| $, the direction of the vector $\mu$. A natural flat prior on $\lambda$ 
is the uniform distribution on the $p$-dimensional sphere for $\lambda$ and the scale prior $\pi(\psi)=1/\sqrt{\psi}$, 
leading to a well-behaved marginal likelihood, see 
\cite{berger:philippe:robert:1998} for precise calculations. However if one considers instead the Jeffreys prior on $\mu$, i.e.
$\pi(\mu) = 1$, then the posterior distribution of $\psi$ is a chi-square distribution with $p$ degrees of freedom and 
non-centrality parameter $|\!| x|\!|^2$, which is not a well-behaved posterior. In particular the posterior mean of $\lambda$ is equal 
to $\hat{\psi} = |\!| x|\!|^2/p+1$ and satisfies $\hat{\psi}-\psi \rightarrow 2$ as $p$ goes to infinity. 

The above examples do not imply that one should not use improper priors on nuisance parameters, since in most cases little information 
is known on those parameters. Rather they show that one needs to be quite careful in selecting improper priors in such cases. The 
construction of Bernardo's (\citeyear{bernardo:1979}) reference priors is particularly relevant in such frameworks.

In the following section, we describe Bayesian testing and Bayesian model comparison or model selection. It is to be noted that model selection can be viewed as a specific example of nuisance parameter framework, where the parameter of interest is the model and the nuisance parameters are the parameters in each model.

\section{Testing versus model comparison}\label{seq:test}

\subsection{Bayes factors}\label{sec:BF}
The most standard Bayesian answer to a testing problem for hypotheses written as $H_0:\,\theta\in\Theta_0$ for the null 
and as $H_1:\,\theta\in\Theta_1$ for the alternative, is the Bayesian estimate corresponding to the 0--1 loss function, i.e.~to the procedure
accepting $H_0$ if and only if 
$$
P^\pi\left[ \Theta_0|x\right]> P^\pi\left[\Theta_1|x\right]\,.
$$
In less formal terms, the null hypothesis is accepted if it is more probable under the posterior distribution than 
under the alternative, which is a very intuitive answer. To constrain the impact of the prior probabilities, a 
different quantity is usually adopted, namely the Bayes factor \citep{kass:raftery:1995}, which is defined by 
\cite{jeffreys:1939,jaynes:2003} as
$$
B_{01} = \displaystyle{ \frac{\pi(\Theta_0|x)}{ \pi(\Theta_1|x)} \bigg/
      \frac{\pi(\Theta_0) }{ \pi(\Theta_1)} }
  = \frac{ \displaystyle{\int_{\Theta_0} f(x|\theta) \pi_0(\theta) \text{d}\theta} }{
       \displaystyle{\int_{\Theta_1} f(x|\theta) \pi_1(\theta) \text{d}\theta} }\,.
$$
Note that the posterior odds can be recovered from the Bayes factor by assigning the appropriate prior probabilities on each of both models, 
contradicting  the criticism of  \cite{templeton:2008}  that the Bayes factor is not scaled in probability terms. Interestingly 
$B_{10}=1/B_{01}$, hence there is no asymmetry in the  definition and construction of Bayes factor, contrariwise to the Neyman--Pearson 
approach. We do not believe that this is a drawback and would rather question the interest in forcing such an asymmetry in the 
Neyman--Pearson tests.

The Bayes factor, a monotonic transform of the posterior probability of $H_0$ which eliminates the influence of the prior weight $\pi(\Theta_0)$,
has a similar interpretation to the classical likelihood ratio. As noted  in the previous section, by integrating out the parameters 
within each hypothesis, the uncertainty on each parameter is taken into account, which induces a natural penalisation for richer models, 
as intuited by \cite{jeffreys:1939} ({\em variation is random until the 
contrary is shown; and new parameters in laws, when they are suggested, must be tested one at
a time, unless there is specific reason to the contrary}). Although we strongly dislike using the term 
because of its undeserved weight of academic authority, the Bayes factor acts as a natural {\em Ockham's razor.} The well-known 
connection with the BIC (Bayesian information criterion, see \citealp{robert:2001}, Chapter 5), with a penalty term of the form 
$d\log n/2$, makes explicit the penalisation induced by Bayes factors in regular parametric models. However it goes beyond 
this class of models, and in much greater generality, the Bayes factor corresponds asymptotically to a likelihood ratio 
with a penalty of the form $d^* \log n^* /2$ where $d^*$ and $n^*$ can be viewed as the effective dimension of the model and the effective number 
of observations, respectively, see \citep{berger:ghosh:2003,chambaz:rousseau:2008}. The Bayes factor therefore offers the major interest 
that it does not require to compute a complexity measure (or penalty term)---in other words, to define what is $d^*$ and what is  $n^*$---, 
which often is quite complicated and may depend on the true distribution.

\subsection{Difficulties}\label{sec:diff}
The inferential problems of Bayesian model selection and of Bayesian testing are clearly those for which the most vivid criticisms can
be found in the literature: witness \cite{senn:2008} who states that {\em the Jeffreys-subjective synthesis betrays a much 
more dangerous confusion than the Neyman-Pearson-Fisher synthesis as regards hypothesis tests}. We find this suspicion rather
intriguing given that the Bayesian approach is the only one giving a proper meaning to the probability of a null hypothesis,
$\mathbb{P}(H_0|x)$, since alternative methodologies can at best specify a probability value on the {\em sampling} space, i.e.~on
the wrong dual space.

If we consider the special case of point null hypotheses---which is not such limited a scope since it includes all variable
selection setups---, there is a difficulty with using a standard prior modelling in this environment. As put 
by \cite{jeffreys:1939}, when {\em considering whether a location parameter $\alpha$ is $0$ [when] the prior is uniform, we 
should have to take $f(\alpha)=0$ and $B_{10}$ would always be infinite}. This is therefore a case when the inferential question 
implies a modification of the prior, justified by the information contained in the question. While avoiding the whole issue is a solution, 
as with Gelman (\citeyear{gelman:2008}) having {\em no patience for statistical methods that assign positive probability to point 
hypotheses of the $\theta = 0$ type that can never actually be true}, considering the null and the alternative hypotheses as
two different models allows for a Bayes factor representation and corresponds to assigning a positive probability to the null hypothesis. 

In our view, one of the major drawbacks of Bayes factors - or even posterior odds - is that they cannot be used under improper priors, 
for lack of proper normalising constants. This is even more acute a difficulty than what is described in Section \ref{sec:nuisance}, because the Bayes factor is simply not defined under improper priors, for any sample size. Solutions have been proposed, akin to cross-validation techniques
in the classical domain \citep{berger:pericchi:1996,berger:pericchi:varshavsky:1998}, but they are somehow too ad-hoc to convince the
entire community (and obviously beyond). In some situations, when parameters shared by both models have the same meaning in each of the 
models, an improper prior can be used on these parameters, in both models.

\newcommand\by{\mathbf{y}}
\newcommand\bX{\mathbf{X}}
For instance, when considering variable selection in a regression model,
$$
\by|\bX,\beta,\sigma \sim \mathcal{N}(\bX\beta,\sigma^2 I_n)\,,
$$
e.g.~when deciding whether or not the null hypothesis $H_0:\beta_1=0$ holds, the 
relevant non informative prior distribution is Zellner's (\citeyear{zellner:1986}) $g$-prior,
where $\pi(\beta|\sigma)$ corresponds to a normal $\mathcal{N}(0,n\sigma^2(\bX^\text{T}\bX)^{-1})$ distribution on 
$\beta$ and a ``marginal" improper prior on $\sigma^2$, $\pi(\sigma^2) = \sigma^{-2}$. This means that, when considering
the submodel corresponding to the null hypothesis $H_0:\beta_1=0$, with parameters $\beta^{(-1)}$ and $\sigma$, we can also
use the ``same" $g$-prior distribution  
$$
\beta^{(-1)}|\sigma,\bX \sim \mathcal{N}(0,n\sigma^2(\bX_{-1}^\text{T}\bX_{-1})^{-1})\,,
$$
where $\bX_{-1}$ denotes the regression matrix missing the column corresponding to the first regressor, and
$\sigma^2\sim\pi(\sigma^2) = \sigma^{-2}$. Since $\sigma$ is a nuisance parameter in this case, we may use the 
improper prior on $\sigma^2$ as {\em common} to all submodels and thus avoid the indeterminacy in the normalising factor of the
prior when computing the Bayes factor
$$
B_{01} = \dfrac{\int f(\by|\beta_{-1},\sigma,\bX) \pi(\beta^{(-1)}|\sigma,\bX_1) \frac{\text{d}\beta_{-1}\,\text{d}\sigma}{\sigma^{2}}}
{\int f(\by|\beta,\sigma,\bX) \pi(\beta|\sigma,\bX) \frac{\text{d}\beta\,\text{d}\sigma}{\sigma^2}}
$$
Figure \ref{fig:from:Core} reproduces an output from \cite{marin:robert:2007} that illustrates how this default prior and the corresponding
Bayes factors can be used in the same spirit as significance levels in a standard regression model, each Bayes factor being associated with 
the test of the nullity of the corresponding regression coefficient. For instance, only the intercept and the coefficients of $X_1,X_2,X_4,X_5$
are significant. This output mimics the standard {\sf lm R} function outcome in order to 
show that the level of information provided by the Bayesian analysis goes beyond the classical output, not to show that we can get 
similar answers to those of a least square analysis since, else, {\em if the Bayes estimator has
good frequency behaviour then we might as well use the frequentist method} \citep{wasserman:2008}. (While computing issues are
addressed in the following Chapter, we stress that all items in the table of Figure \ref{fig:from:Core} are obtained via
closed form formulae.)

\begin{figure}
{\sffamily

\begin{tabular}{l l l l}
            &Estimate  &BF        &log10(BF)\\
& & & \\
(Intercept)  &9.2714   &26.334  &1.4205 (***) \\
X1          &-0.0037   &7.0839  &0.8502 (**) \\
X2          &-0.0454   &3.6850  &0.5664 (**) \\
X3          &0.0573    &0.4356  &-0.3609 \\
X4          &-1.0905   &2.8314  & 0.4520 (*) \\
X5          & 0.1953   &2.5157  & 0.4007 (*) \\
X6          &-0.3008   &0.3621  &-0.4412 \\
X7          &-0.2002   &0.3627  &-0.4404 \\
X8          & 0.1526   &0.4589  &-0.3383 \\
X9          &-1.0835   &0.9069  &-0.0424 \\
X10         &-0.3651   &0.4132  &-0.3838 \\
\end{tabular}

\medskip
evidence against H0: (****) decisive, (***) strong,
(**) substantial, (*) poor
}

\caption{\label{fig:from:Core} {\sf R} output of a Bayesian regression analysis on a processionary caterpillar
dataset with ten covariates analysed in \cite{marin:robert:2007}. The Bayes factor on each row corresponds to
the test of the nullity of the corresponding regression coefficient.}
\end{figure}

The major criticism addressed to the Bayesian approach to testing is therefore that it is not interpretable on the same scale as
the Neyman-Pearson-Fisher solution, namely in terms of probability of Type I error and of power of the tests. In other words, 
{\em frequentist methods have coverage guarantees; Bayesian methods don't; 95 percent frequentist intervals will live up to their 
advertised coverage claims} \citep{wasserman:2008}. A natural question is then to question the appeal of such frequentist
properties when considering a single dataset, i.e.~in Jeffreys' (1939) famous words, {\em a hypothesis that may be true
may be rejected because it had not predicted observable results that have not occurred}, especially when considering that $p$-values
may be inadmissible estimators \citep{hwang:casella:robert:wells:farrel:1992}. From a decisional perspective---with which the
frequentist properties should relate---, a classical Neyman-Pearson-Fisher procedure is never evaluated in terms of the consequences
of rejecting the null hypothesis, even though the rejection must imply a subsequent action towards the choice of an alternative model.
Therefore, complaining that {\em having a high relative probability does not mean that a hypothesis is true or supported by the data}
\citep{templeton:2008}, simply because the Bayesian approach is relative in that it {\em posits two or more alternative hypotheses and 
tests their relative fits to some observed statistics} \citep{templeton:2008}, is missing the main purpose of tests, which is not to 
validate or invalidate a golden model {\em per se} but rather to infer a working model that allows for acceptable predictive 
properties.\footnote{It is worth repeating the earlier assertion that all models are false and that finding that a hypothesis 
is ``true" is not within our reach, if at all meaningful!}

\subsection{Model choice}\label{sec:mocho}
For model choice, i.e.~when several models are under comparison for the same observation
$$
\mathfrak{ M}_i : x \sim f_i(x|\theta_i)\,, \qquad i \in \mathfrak{I}\,,
$$
where $\mathfrak{I}$ can be finite or infinite, the usual Bayesian answer is similar to the Bayesian tests as described above. 
The most coherent perspective (from our viewpoint) is actually to envision the tests of hypotheses as particular cases of model
choices, rather than trying to justify the modification of the prior distribution criticised by \cite{gelman:2008}. This also also
to incorporate within model choice the alternative solution of model averaging, proposed by \cite{madigan:raftery:1994}, which
strives to keep all possible models when drawing inference.

The idea behind Bayesian model choice is to construct an overall probability on the collection of models 
$\cup_{i \in \mathfrak{I}} \mathfrak{M}_i$ in the following way: the parameter is $\theta = (i, \theta_i)$, i.e. the model index and given the model index equal to $i$, the parameter $\theta_i$ in model $\mathfrak{M}_i$, then the prior measure on the parameter $\theta$ is expressed as
$$
\text{d}\pi(\theta) = \sum_{i\in \mathfrak{I}} p_i \text{d}\pi_i(\theta_i), \quad \sum_{i\in I_i} p_i = 1\,,
$$
where both the $\pi_i$'s and $p_i$'s are part of the prior modelling, hence chosen by the experimenter. (The $\pi_i$'s have the
natural interpretation of the traditional prior under model $\mathfrak{M}_i$, while the $p_i$'s correspond to the prior assessment
of the models under comparison.) As a consequence, the Bayesian model selection associated with the 0--1 loss function  
and the above prior is the model that maximises the posterior probability
$$
\pi({\mathfrak M}_i|x) = \dfrac{ \displaystyle{p_i \int_{\Theta_i} f_i(x|\theta_i)
         \pi_i(\theta_i) \text{d}\theta_i}}{
\displaystyle{\sum_j p_j \int_{\Theta_j} f_j(x|\theta_j) \pi_j(\theta_j)
      \text{d}\theta_j}  }
$$
across all models.  Contrary to classical pluggin likelihoods, the marginal likelihoods involved in the above ratio 
do compare on the same scale and do not require the models to be nested: the criticism that {\em complicating dimensionality 
of test statistics is the fact that the models are often not nested, and one model may contain parameters that do not have 
analogues in the other models and vice versa}
\citep{templeton:2008} is not founded. As mentioned in Section \ref{sec:nuisance} integrating 
out the parameters $\theta_i$ in each of the models takes into account their uncertainty thus the marginal likelihoods 
$ \int_{\Theta_i} f_i(x|\theta_i) \pi_i(\theta_i) \text{d}\theta_i$ are naturally penalised likelihoods. In many setups, 
the Bayesian model selector as defined above is consistent, i.e. as the number of observations increases the probability 
of choosing the right model goes to 1.

\subsection{Other issues}\label{sec:oza}
The computational requirements related to handling a collection of marginal likelihoods will be addressed 
in the following Chapter, in connection with the review of classical solutions in \cite{robert:marin:2010}.
Interestingly enough, the most accurate approximation technique for marginal likelihoods
is, when applicable, directly derived from Bayes theorem, via Chib's (\citeyear{chib:1995}) rendering:
$$
m(x) = \dfrac{\pi(\theta) f(x|\theta)}{\pi(\theta|x)} \approx \dfrac{\pi(\theta) f(x|\theta)}{\hat\pi(\theta|x)}\,, 
$$
where $\hat\pi(\theta|x)$ is a simulation-based approximation to the posterior density based on simulated latent variables.
\cite{marin:robert:2008} illustrate this method in the setting of mixtures and \cite{robert:marin:2010} in the 
alternative case of a probit model, respectively, both of which demonstrate the precision of this approximation.\footnote{There 
have been discussions about the accuracy of this method in multimodal settings \citep{fruhwirth:2004}, but straightforward 
modifications \citep{berkhof:mechelen:gelman:2003,lee:marin:mengersen:robert:2008} overcome such difficulties and make for both 
an easy and a well-grounded computational tool associated with Bayes factors.}

Posterior odds and Bayes factors are the most common Bayesian  approaches to testing, however they are not the only ones. 
In particular the choice of the 0--1 loss function is not necessarily relevant or the most relevant. In some situations it 
might be more interesting to penalise the loss with the distance to the null hypothesis for instance, 
see \cite{robert:rousseau:2002,rousseau:2007} where such ideas are applied to goodness of fit tests 
or \cite{bernardo:2009}.

\section{On pervasive computing}
Bayesian analysis has long been derided for providing optimal answers that could not be computed. With the advent of
early Monte Carlo methods, of personal computers, and, more recently, of more powerful Monte Carlo methods \citep{hitchcock:2003}, the
pendulum appears to have switched to the other extreme and {\em Bayesian methods seem to quickly move to elaborate computation}
\citep{gelman:2008}, a feature that does not make them less suspicious: {\em a simulation method of inference hides unrealistic 
assumptions} \citep{templeton:2008}. The simulation techniques that have done so much to
promote Bayesian analysis in the past decades are detailed in the next Chapter and thus not described here. 
We nonetheless want to point out that, while simulation methods can be misused---as about
any other methodology---and while {\em Bayesian simulation seems stuck in an infinite regress of inferential uncertainty} 
\citep{gelman:2008}, there exist enough convergence assessment techniques \citep{robert:casella:2009} to ensure a reasonable confidence 
about the approximation provided by those simulation methods. Thus, as rightly stressed by \cite{bernardo:2008}, {\em the discussion of 
computational issues should not be allowed to obscure the need for further analysis of inferential questions}.\footnote{The confusion of
\cite{templeton:2008} is of this nature, addressing the principles of Bayesian inference when aiming at the ABC simulation methodology.}

The field of Bayesian computing is therefore very much alive and, while its diversity can be construed as a drawback
by some, we do see the emergence of new computing methods adapted to specific applications as most promising, because it bears 
witness to the growing involvement of new communities of researchers in Bayesian advances.


\begin{theacknowledgments}
C.P.~Robert and Judith Rousseau are both supported by the 2007--2010 ANR-07-BLAN-0237-01 ``SP Bayes" grant.
\end{theacknowledgments}

\small
\bibliographystyle{ims} 

\end{document}